\documentclass[preprint,showpacs,aps,a4paper]{revtex4}

\usepackage{graphicx}
\usepackage{dcolumn}
\usepackage{bm}
\usepackage[english,spanish,activeacute]{babel}
\def\beq{\begin{equation}}
\def\eeq{\end{equation}}
\def\nsec{N_{\rm sec}}
\def\ncol{N_{\rm coll}}
\def\elead{E_{\rm lead}}
\def\eprim{E_{\rm prim}}
\def\esec{E_{\rm sec}}
\def\kinel{k_{\rm inel}}
\def\xmax{X_{\rm max}}
\def\gcm#12{g/cm$^2$}

\begin{document}
\selectlanguage{english}
\title{Influence of diffractive interactions on cosmic ray air
  showers}

\author{R. Luna\footnote{Also at ESCOM-IPN, M\'exico DF, M\'exico.}
and A. Zepeda}
\affiliation{
Departamento de F\'{\i}sica, Cinvestav,
Av.~IPN 2508, Col.~San Pedro Zacatenco,
07360 M\'exico DF, M\'exico.
}
\author{C. A. Garc\'{\i}a Canal and S. J. Sciutto}
\affiliation{
Departamento de F\'{\i}sica and IFLP/CONICET,
Universidad Nacional de La Plata,
C. C. 67 - 1900 La Plata,
Argentina.
}

\begin{abstract}
A comparative study of commonly used hadronic collision simulation
packages is presented. The characteristics of the products of
hadron-nucleus collisions are analyzed from a general perspective, but
focusing on their correlation with diffractive processes.
One of the purposes of our work is to give quantitative
estimations of the impact that different characteristics of the hadronic
models have on air shower observables. Several sets of shower
simulations using different settings for the parameters controlling
the diffractive processes are used to analyze the correlations between
diffractivity and shower observables. We find that the relative
probability of diffractive processes during the shower development
have a non negligible influence over the longitudinal profile as well
as the distribution of muons at ground level. The implications on
experimental data analysis are discussed.
\end{abstract}

\maketitle

\section{Introduction}

The measurement of extensive air showers (EAS) is presently the way to
study cosmic rays with primary energy above several hundreds of TeV. The
properties of primary cosmic rays have to be deduced from the development
of the shower in the atmosphere, and from the characteristics of the
secondaries detected at the observation level.

Due to the lack of experimental data on particle interactions at the
highest energies, it is necessary to interpret EAS measurements by
comparing them with model predictions. Due to the complexity of the
interactions that take place during the shower development, such
predictions are generally obtained from Monte Carlo simulations.

The algorithms used to perform such simulations include sections that
correspond to the different interactions that the secondary particles
can undergo during their propagation. Among them, the hadronic
interactions are one of the most difficult to model accurately, while
playing a key role when trying to predict the final observables of a
shower \cite{augersimap}.  Such models are affected by uncertainties
that cannot, at present, be totally controlled \cite{EngelReview}. The
uncertainties come from approximations that are intrinsic to the
respective models, plus uncertainties and inconsistencies in the
experimental data used to calibrate model parameters, plus the
uncertainties associated with extrapolations outside the range covered
by the available experimental data.

An important task is, therefore, to give quantitative estimation of
the impact of these uncertainties on shower observables and on the
estimations of properties of the primary particle. This question is of
particular importance in ultra high energy cosmic ray observatories
such as the Pierre Auger and even more relevant for primary energy
determination from fluorescence emitance since this effect depends on
the fraction of the cascade that produces the greatest portion of
fluorescence light and in turn this fraction depends on the hadronic
model.

The aim of the present work is aligned in this direction: well-known
packages like SIBYLL \cite{SIBYLL21,SIBYLL21_2,SIBYLL21_3}, QGSJET
\cite{QGSJET01,QGSJET01_2}, and DPMJET
\cite{DPMJET2,DPMJET2_1,DPMJET2_2,DPMJET2_3} are extensively
compared. Our study is carried out with a very practical approach,
analyzing first the secondaries produced after individual collisions,
and then measuring the impact in the shower development of different
hadronic configurations. This work covers several aspects of the
hadronic interactions, but it is particularly concentrated in the
study of the diffractive interactions. The analysis here presented is
complementary to a work reported in references
\cite{augersimap,PlayaDelCarmen}.

In section \ref{sec:hc0} hadronic collisions, and in particular
diffractive ones, are described. In section \ref{sec:cahm} we compare the
properties of DPMJET, SYBILL and QGSJET hadronic models in collisions
of protons with air with energy from $10^{2}$ to $10^{9}$ GeV. In
section \ref{sec:eso} we compare the impact that the different alternatives for
modeling the diffractive hadronic interactions have on common air
shower observables. In section \ref{sec:conclu} we present our
final remarks and conclusions.

\section{Hadronic collisions}
\label{sec:hc0}

From a practical point of view, a hadronic collision can be described
as a process where an incident particle $P$, called the projectile,
interacts with a target $A$ --normally a nucleus of $A=Z+N$ nucleons
($Z$ protons and $N$ neutrons)-- to produce $\nsec$ secondary
particles $S_1,\ldots,S_{\nsec}$.  Such secondary particles are
generally hadrons, but can eventually be nuclear fragments, photons,
etc., depending on the characteristics of the collision.  In those
cases where the primary particle survives after the collision (with
changed energy and momentum), the surviving primary is accounted just
as one more secondary particle within the $S_1,\ldots,S_{\nsec}$ set.

Each secondary $S_i$ ($i=1,\ldots,\nsec$) is characterized by its
particle type, its energy $E_{S_i}$, and the angle $\theta_i$ between
the primary and secondary momenta. It is also possible to define an
azimuthal angle for each secondary. To adequately quantify the
directions of the secondary particles it is convenient to use the
so-called {\em pseudorapidity,\/} defined as
$
\eta = - \ln\tan(\theta/2)
$, 
instead of specifying the angle $\theta$ directly.

Let $\elead$ be the energy of the secondary with maximum energy (the
so-called leading particle). The {\em leading energy
fraction,\/} $f_L$, for the collision, is defined as
\beq                            \label{eq:fldef}
f_L = \frac{\elead}{E_P} .
\eeq
For high energy primaries the energies of the
secondaries satisfy
$
\sum_{i=1}^{\nsec} E_{S_i} \le  E_P
$
and, as a consequence, one has $0<f_L < 1$.
The {\em inelasticity,\/} $\kinel$, is defined as the fraction of
energy carried by all the secondary particles, excluding the leading
secondary,
\beq                              \label{eq:ineldef}
\kinel = 1 - f_L .
\eeq

In normal hadronic collisions $f_L$ or, equivalently, the
inelasticity, fluctuates virtually in
all the allowed range from 0 to 1. In one extreme one has the hard
interactions with large momentum transfers producing many secondaries
that share the available primary energy. On the other hand,
the so-called {\em diffractive dissociation events,\/}
are characterized by low multiplicity and fast secondary particles,
that imply $f_L$ close to 1.

The diffractive interactions play a very important role during the
development of air showers, due to the fact that they provide a way 
of 
transporting substantial amounts of energy deep in the atmosphere, and
turn into a critical factor that controls the global characteristics of the
shower profile \cite{augersimap}.

The results coming from different theoretical treatments of soft
interactions are not always coincident; and they cannot be
conclusively checked against experimental data because up to the
present time these forward processes could not be measured with enough
accuracy in collider experiments \cite{EngelReview,PlayaDelCarmen}.

For all these reasons we consider important to investigate the
properties of different quantities associated with hadronic
interactions or air shower development, taking into account explicitly
the diffractive or non-diffractive nature of the corresponding
hadronic processes.

As mentioned before, in a diffractive collision there is a leading
particle whose energy is clearly larger than the energies of the other
secondaries ($f_L$ close to 1). Additionally, the total number of
secondaries is generally small. On the other hand, a properly
inelastic collision at very high energies is characterized by a large
number of secondaries of comparable energy ($f_L \ll 1$).  We can
therefore make simultaneous use of $\nsec$ and $f_L$, or equivalently
$\kinel$, to distinguish diffractive from inelastic collisions. This
approach proves to work acceptably in practice, having the advantage
of being applicable to every collision generator (and even to the
analysis of experimental data) where usually there is no additional
information that permits discriminating the cases of true diffractive
processes from the other non-diffractive ones.

\section{Comparative analysis of hadronic models}
\label{sec:cahm}

As a first step in our study, we have performed
a comparative analysis of the output coming from different hadronic
packages when running them with a common input.

We have run batches of $\ncol$ events ($\ncol = 10,000$ unless
otherwise specified) for each combination of primary type, primary
energy, and hadronic package. After each call to the hadronic
procedures, a list of secondaries was obtained, with short-lived
products (resonances) forced to decay. These secondaries were then
processed to identify the leading particle, and to plot in histograms
the quantities introduced in section \ref{sec:hc0}. Finally $f_L$ was
evaluated, and analyzed in combination with the number of secondaries
and the properties of the leading particle in order to label the
collisions as ``diffractive'' or ``non-diffractive''.  We have
included in our analysis three of the most popular high energy
interaction models, namely SIBYLL 2.1 \cite{SIBYLL21},
QGSJET01\cite{QGSJET01}, and DPMJET 2.5 \cite{DPMJET2}.

Every one of these models is capable of processing hadronic collisions
having a hadron as projectile, and a specified nucleus as target. The
energy of the particle must lie in a determined range, characteristic
of each model. In the present work these ranges have been
taken as $E_P \ge 30$ GeV for DPMJET and QGSJET, and $E_P\ge 100$ GeV
for SIBYLL. With respect to the targets, we use a mixture of
nitrogen and oxygen to emulate collisions in air, the medium of
propagation of cosmic ray air showers.

\subsection{Multiplicity and inelasticity}
\label{sec:mandi}

To start with our analysis of individual collisions, let us consider
first the number of secondaries.

\begin{figure}[tp] 
\begin{center}
\includegraphics[width=4.5cm]{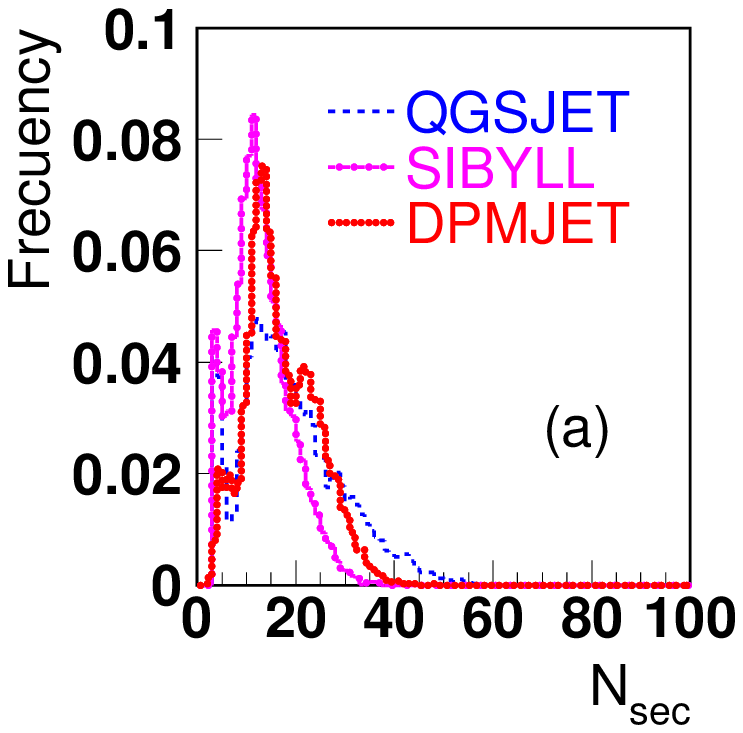}
\includegraphics[width=4.5cm]{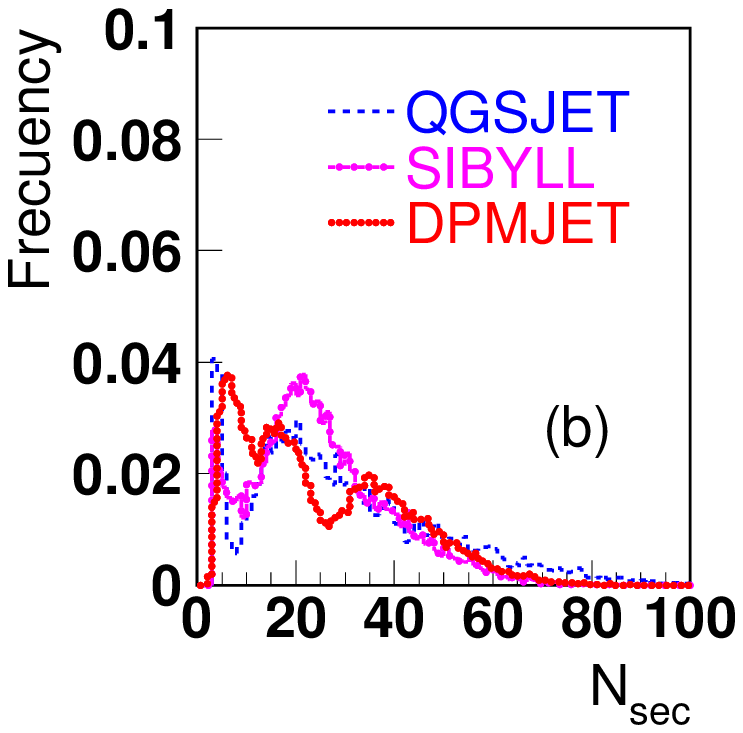}
\includegraphics[width=4.5cm]{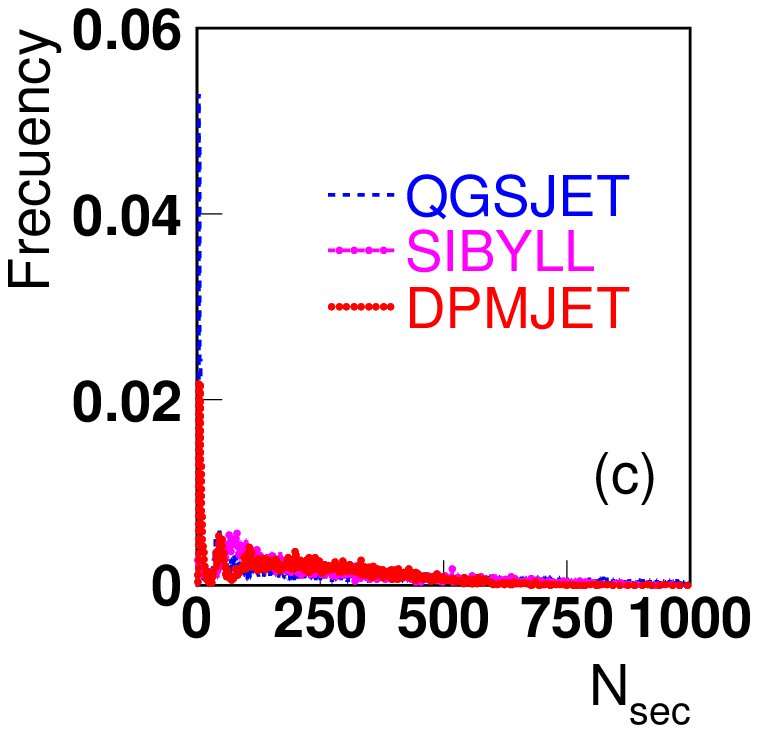}
\end{center}
\caption{$\nsec$ distributions for proton-air collisions at 100 GeV
  (a), 1 TeV (b), and 100 PeV (c).}
\label{fig:nsecdist}
\end{figure}

\begin{figure}[tp]
\begin{center}
\includegraphics{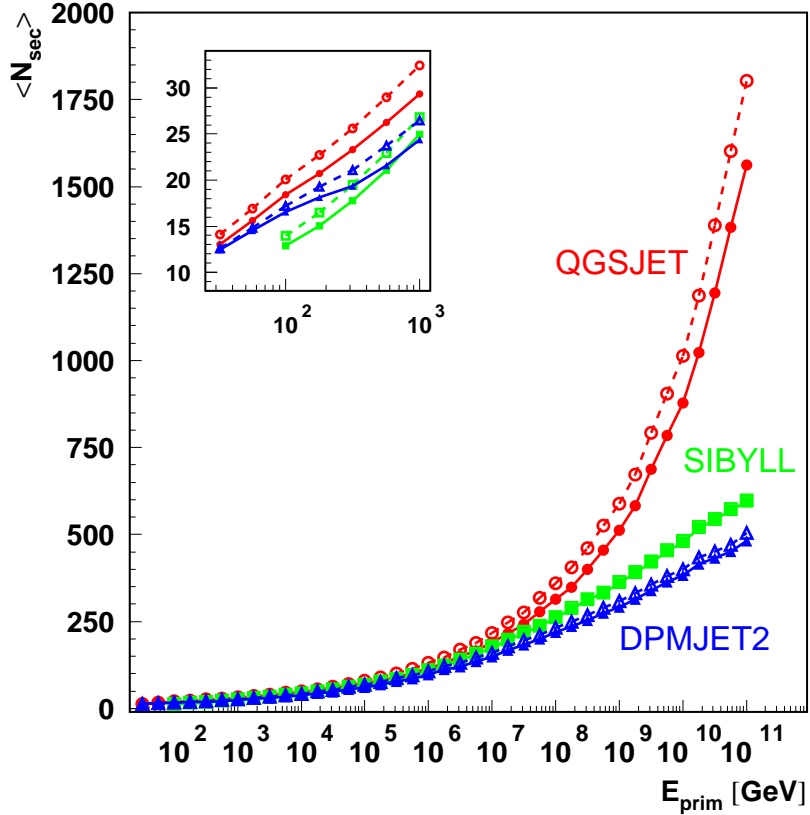}
\end{center}
\caption{Average number of secondaries in proton-air collisions versus
  primary energy. The solid (open) symbols correspond to averages over all
  (non-diffractive) events. The lines are only to guide the eye. The
  low energy region is shown in more detail in the inset.
}
\label{fig:nsecvseproton}
\end{figure}

In figure \ref{fig:nsecdist}, distributions of numbers of secondaries are
displayed for several representative primary energies. The diffractive
interactions show up clearly at each plot as a characteristic peak in
the few-secondary zone of the abscissas.
We can see that there are evident differences among the plots
corresponding to different models, especially when comparing QGSJET
with the other models.
An outstanding feature is the well known fact that QGSJET produces
substantially more secondaries than SIBYLL or DPMJET, especially at
very high energies \cite{PlayaDelCarmen,doqui}. This fact shows up
clearly in figure \ref{fig:nsecvseproton} where the average number of
secondaries is plotted versus the primary energy. The curves with
solid lines and symbols correspond to averages considering all kinds
of events, while the curves with dashed lines and open symbols
correspond to averages over non-diffractive events only.

The general averages are always smaller than the ones over
non-diffractive events, as expected, since diffractive events have
very few secondaries and therefore tend to reduce averages when
included in the samples.

The differences between general and non-diffractive cases are
significant in the case of QGSJET, small in the case of DPMJET, and
almost negligible in the case of SIBYLL. A
similar behaviour can be found in the case of
pion primaries (plots not included for brevity).

\begin{figure}[tp]
\begin{center}
\includegraphics{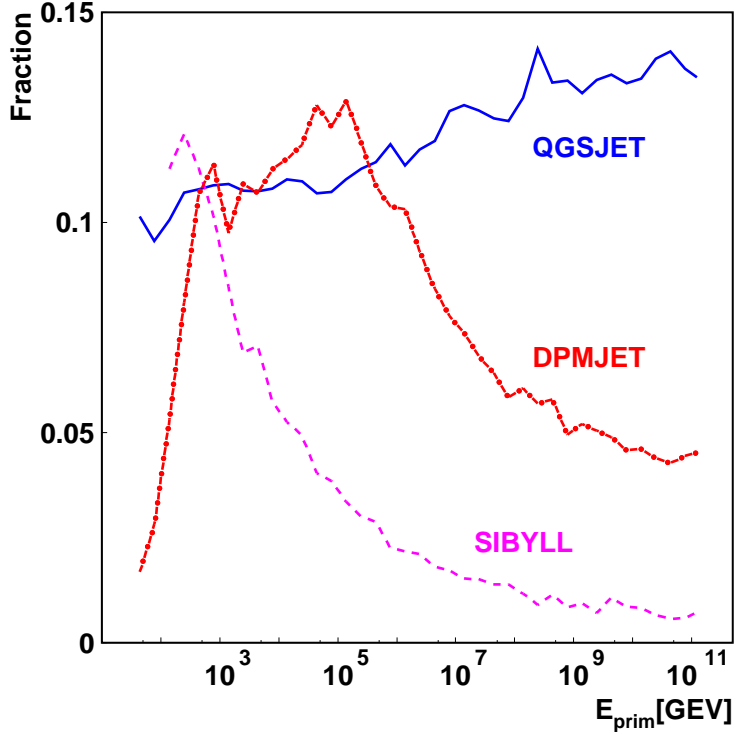}
\end{center}
\caption{Fraction of diffractive events versus primary energies for
  the case of proton-air collisions.}
\label{fig:diffraction}
\end{figure}

The main characteristics of the preceding plot are better
understood considering that the influence of diffractive events in a
sample is not only due to the properties of the diffractive
interaction itself, but also to the magnitude of their relative
probability. In figure \ref{fig:diffraction}, the fractions of
diffractive events registered in our runs is plotted as a function of
primary energy, in the case of proton primaries. The very significant
difference between the QGSJET and SIBYLL cases is one of the
outstanding features of this plot: these results indicate that in
QGSJET the ratio between the diffractive and total cross sections
does not suffer substantial variations in the whole range of energies
considered (from 30 GeV to 100 EeV), while the corresponding
cross section ratio for SIBYLL presents a completely different
behavior, decreasing as the primary energy increases. This explains
clearly why the SIBYLL and DPMJET curves in figure \ref{fig:nsecvseproton} 
virtually overlap at the highest
energies. Notice also that the relatively high diffractive probability
of QGSJET tends to compensate the very large number of secondaries
produced by this model in non diffractive interactions.

In between of these two completely different behaviors we can place
the DPMJET case, characterized by a diffractive probability similar
to QGSJET, for primary energies up to $10^{15}$ eV approximately, and
then decreasing continuously for larger primary energies.

It is important to mention that in the case of pion collisions, the
probabilities of diffractive interactions in the cases considered (not
plotted here for brevity) are
very similar to the respective ones for protons.

The SIBYLL and QGSJET curves in figure \ref{fig:diffraction} can be
understood analysing the energy dependence of the diffractive,
$\sigma_{\rm diff}$, and total, $\sigma_{\rm tot} = \sigma_{\rm diff}
+ \sigma_{\rm inel}$, cross sections, and taking into account that the
fractions of diffractive events, $F_{\rm diff}$, plotted at the
mentioned figure are approximately equal to the respective diffractive
to total cross section ratios, that is,
\beq                                     \label{eq:fdiffratio}
 F_{\rm diff} \approx \frac{\sigma_{\rm diff}}{\sigma_{\rm tot}} .
\eeq

Let us discuss first QGSJET. In this model cross sections are calculated on
the basis of the quasieikonal approximation
 \cite{QGSJET01,QGS,QGS_3,Zoller88}, and
it is found that
\begin{eqnarray}
\sigma_{\rm tot} &\propto& \ln^{2}\left(s\right),
                                     \label{eq:sigmatotqgsib}  \\
\sigma_{\rm inel} &\propto&  \ln^{2}\left(s\right),
                                     \label{eq:sigmainelqgsib}  \\
\sigma_{\rm diff} &\propto& \ln^{2}\left(s\right) ,
                                     \label{eq:sigmadiffqgsib}
\end{eqnarray}
where $s$ is the (square of the) center of mass energy.
Then, from equation (\ref{eq:fdiffratio}) it follows that in the
QGSJET case one has $F_{\rm diff} \approx \hbox{const.}$ when
$s\to\infty$, in accordance with the corresponding curve in figure
\ref{fig:diffraction}.

On the other hand, for the SIBYLL case
\cite{augersimap,SIBYLL21_3,Gaisser89,Fletcher92} and in the high
energy limit, the total and inelastic cross sections behave again
proportional to $\ln^2(s)$ as in equations (\ref{eq:sigmatotqgsib})
and (\ref{eq:sigmainelqgsib}), but the diffractive cross section grows
logarithmically, that is
\beq                                          \label{eq:sigmadiffsibyll}
\sigma_{\rm diff}\propto \ln\left(s\right).
\eeq
Then, from equations (\ref{eq:fdiffratio}), (\ref{eq:sigmatotqgsib}),
and (\ref{eq:sigmadiffsibyll}), it is straightforward to see that
$F_{\rm diff} \to 0$ when
$s\to\infty$, similarly as in the SIBYLL curve displayed in figure
\ref{fig:diffraction}.

It should also be noticed that the experimental diffractive cross
section presents a strong saturation effect starting around $\sqrt{s}=
50$ GeV (see for example figure 1 of reference \cite{goulianos}). This
effect is probably related to unitarization. The saturation in
diffraction is in contrast with the growing in energy of the total
cross section \cite{goulianos}. Even without entering a more
detailed theoretical discussion, one can expect that the diffractive
component of the cross section looses protagonism with energy, leading
in turn to a relative fraction of diffractive events that decreases
with primary energy, corresponding qualitatively to the SIBYLL or
DPMJET cases plotted in figure \ref{fig:diffraction}.

\begin{figure}[tp]
\begin{center}
\includegraphics{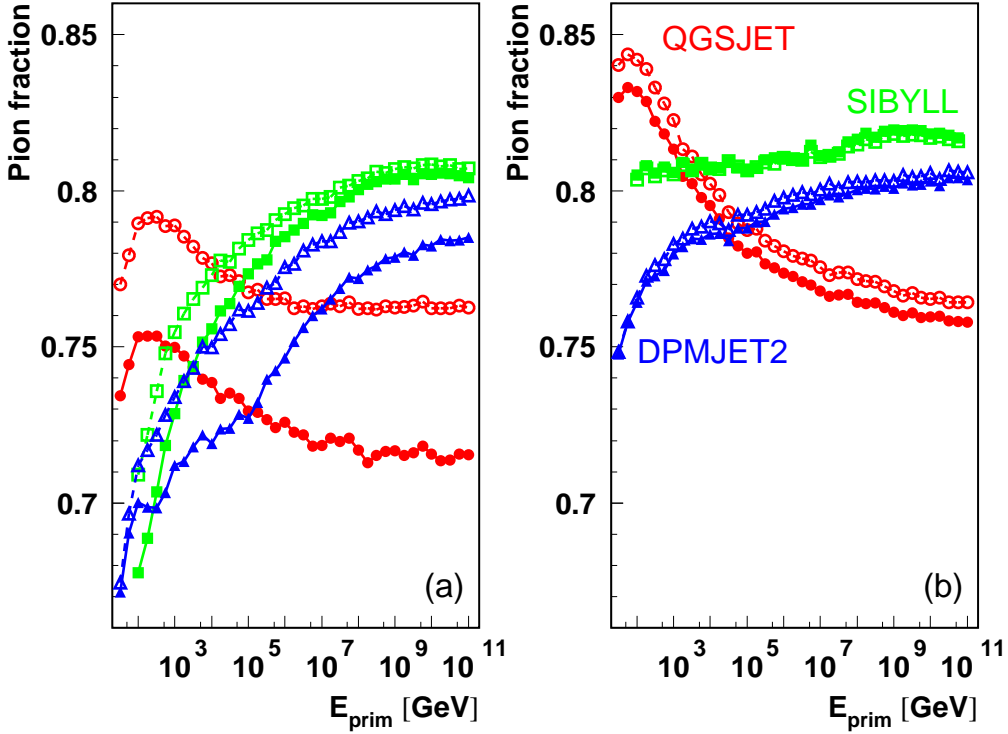}
\end{center}
\caption{Average fraction of pions produced in hadronic collisions
  versus primary energy, in the cases of proton-air (a) and pion-air
  (b) collisions. The solid (open) symbols correspond to averages over all
  (non-diffractive) events.}
\label{fig:pifracvsegy}
\end{figure}

The hadronic models studied here present noticeable differences when
considering the composition of the secondaries generated in nuclear
collisions. A useful quantitative measure of the kind of particles
emerging from such collisions is the fraction of pions, that is,
the total number of pions (charged and neutral) divided by the total
number of secondaries.

In figure \ref{fig:pifracvsegy} the fractions of pions for proton-air
(a) and pion-air (b) collisions are plotted as a function of the
primary energy.  The differences between models are evident: The
SIBYLL and DPMJET fractions rise with energy, while the corresponding
one for QGSJET decreases after reaching a maximum at relatively low
$E_P$. Additionally, the largest differences correspond to energies
around and below 1 TeV, a region of particular importance in the case
of air showers due to the existing direct correlation between the low
energy pion production and the number of muons in the shower. It
should be noticed that the discrepancies in the pion fractions coming
from different models are also present at low energies, as reported in
reference \cite{PlayaDelCarmen}.

\subsection{Energy and transverse momentum}

\begin{figure}[tp]
\begin{center}
\includegraphics{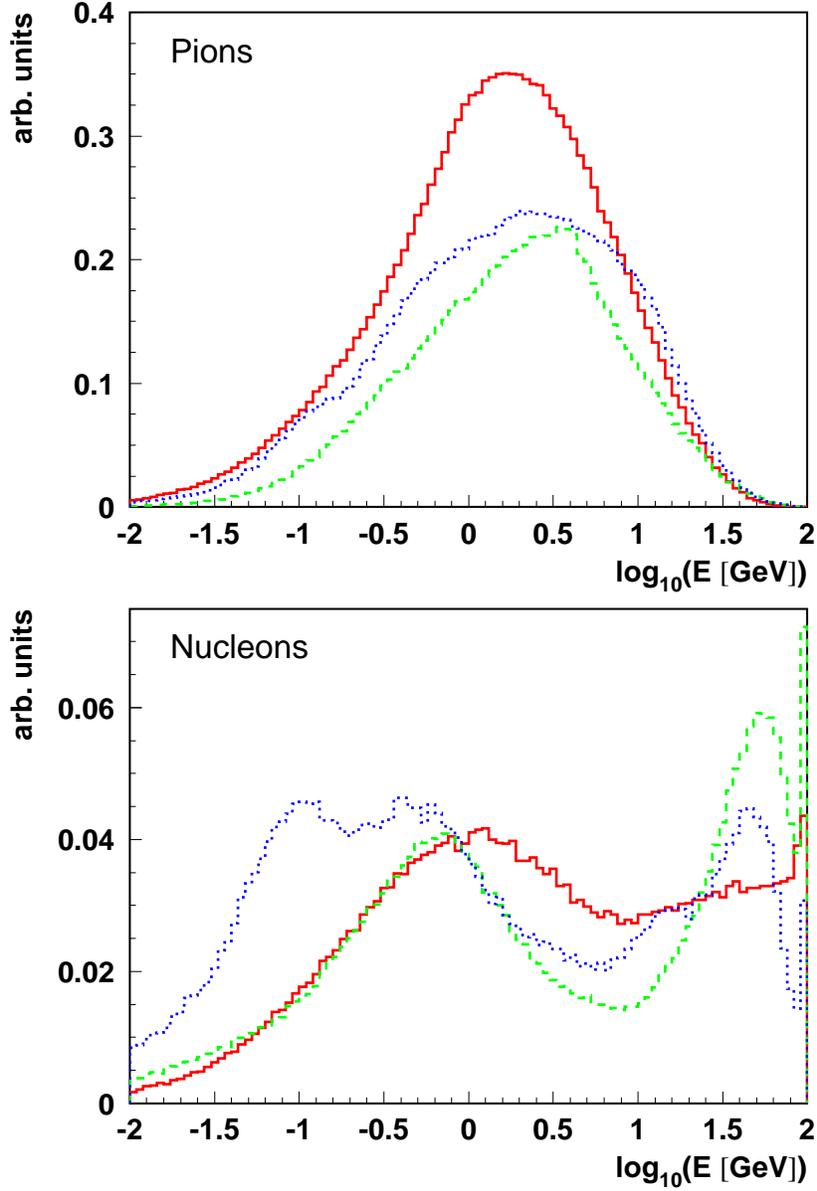}
\end{center}
\caption{Energy distributions of secondary pions and nucleons in the
  case of 100 GeV proton-air collisions. The solid, dashed, and dotted
  histograms correspond to QGSJET, SIBYLL, and DPMJET2, respectively.}
\label{fig:secedistproton100gev}
\end{figure}

In a normal hadronic collision producing many secondaries, the energies
of the emerging particles distribute broadly within the range of all
possible energies $E\le E_P$. This fact is illustrated in figure
\ref{fig:secedistproton100gev}, containing the energy distributions of
pions and nucleons, in the very representative case of 100 GeV
proton-air collisions.

The pion energy distributions are unimodal, and are centered at
energies around 2 to 5 GeV. Notice the larger area in the QGSJET case,
indicating that the average number of pions produced by this model is
larger than the corresponding ones for SIBYLL and DPMJET.

The distributions for nucleons present a more complicated structure,
product of the more involved mechanisms of nucleon production that
enter in action in the different models. These distributions are made
up of two components clearly distinguishable: (i) A neat peak at
$\esec \simeq E_P$, that corresponds to leading nucleons emerging from
diffractive events. (ii) A widely spread distribution that corresponds
to inelastic production of nucleons and antinucleons. This part of the
distribution is in general multimodal, indicating the coexistence of
several production mechanisms with different average secondary energy.

A detailed theoretical explanation of the characteristics of these
distributions is beyond the purpose of the present work. The
interested reader can find further details in the references 
\cite{SIBYLL21,SIBYLL21_2,QGSJET01,QGSJET01_2}.

For meson primaries, the energy distribution of secondary nucleons
(not plotted here for brevity) acquires a simpler structure (much like
the pion energy distribution) with virtually no secondaries having
$\esec \simeq E_P$. This is a clear consequence of the fact that in
this case the diffractive interactions involve mesons as leading
particles.

For larger primary energies, the secondary energy distributions (not
plotted here) maintain approximately most of the features of the
already described distributions at 100 GeV, but extending always in
the entire allowable energy range. In the case of the energy
distribution of pions, the central value increases continuously with
the primary energy.

\begin{figure}[tp]
\begin{center}
\includegraphics{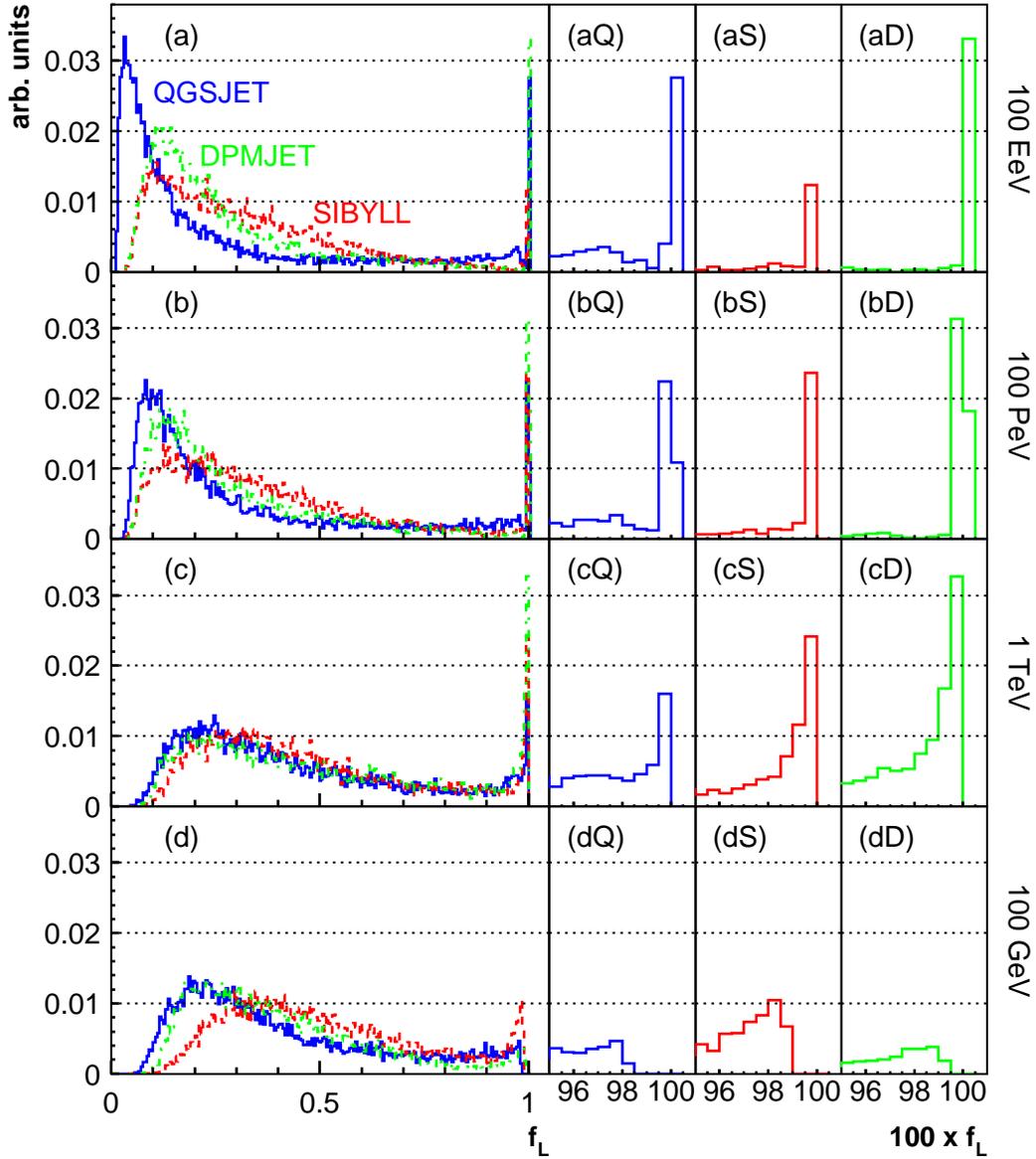}
\end{center}
\caption{Leading energy fraction distributions for proton-air
  collisions at different representative primary energies.}
\label{fig:fldistproton}
\end{figure} 

Another quantity of interest to our analysis is the distribution of
the fraction of energy carried by the leading secondary, $f_L$ (see
the definition at section \ref{sec:hc0}).

Figure \ref{fig:fldistproton} displays typical $f_L$
distributions. The plots include distributions for QGSJET, SIBYLL, and
DPMJET for proton projectiles at several representative energies.

The sharp peaks around $f_L=1$ are the distinctive signature of the
diffractive events (The small plots at the right part of figure
\ref{fig:fldistproton} show these peaks in detail. Notice that the
areas of such peaks are correlated to the respective diffractive
event factions plotted in figure \ref{fig:diffraction}. On the other
hand, in the properly inelastic collisions the available energy is
shared among many secondaries, leading to a fluctuating $f_L$ that
distributes in the whole $[0,1]$ range.

It is worth noticing the particular shape of the QGSJET distribution
at the highest energies (figure \ref{fig:fldistproton} (a)),
that presents two noticeable peaks located at
both $f_L\sim 0$ and $f_L\sim 1$ extremes.
The concentration of events around $f_L\sim 0$ is clearly
more accentuated than in the cases of SIBYLL or DPMJET. This
difference is directly correlated with the very large number of
secondaries produced by QGSJET in inelastic collisions at the
highest energies.

\begin{figure}[tp]
\begin{center}
\includegraphics[width=6.5cm]{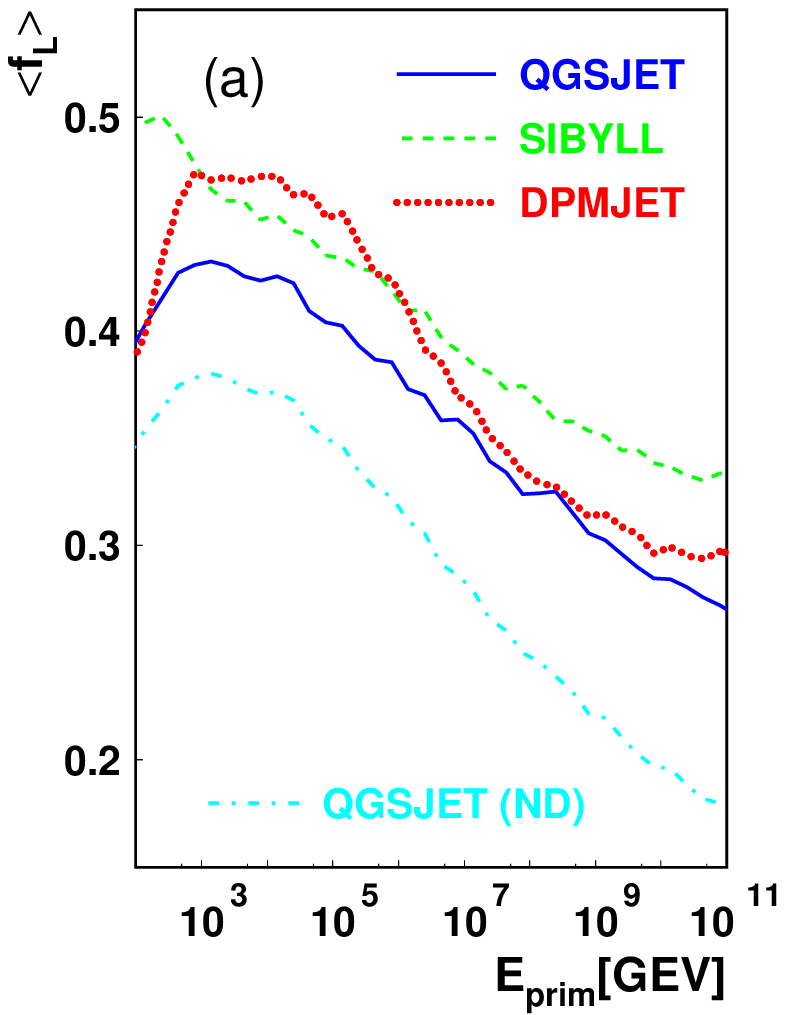}
\includegraphics[width=6.5cm]{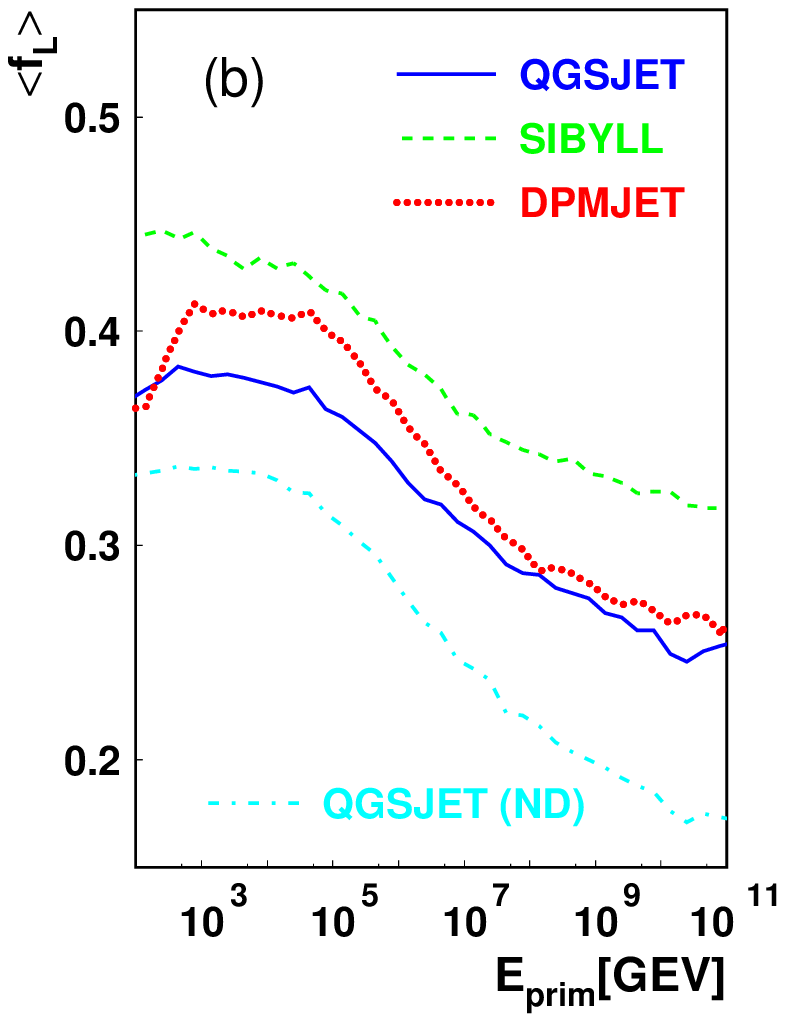}
\end{center}
\caption{$\langle f_L\rangle$ versus $\eprim$ for proton-air (a), and
  pion-air (b) collisions}
\label{fig:flvseprim}
\end{figure}

The plots in figure \ref{fig:flvseprim} also illustrate this
particular characteristic of QGSJET. In this figure the mean $\langle
f_L\rangle$ is plotted as a function of the primary energy for the cases
of proton and pion projectiles. The graphs include two curves for
QGSJET, namely, the general average, and the average excluding
diffractive processes. This last one indicates clearly that the
fraction of energy carried away by the leading particle is sensibly
lower than in every other case, in agreement with the data displayed
in figure \ref{fig:fldistproton} for proton collisions at
representative fixed energies.

Notice also that at the highest energies the largest $\langle f_L
\rangle$ corresponds to SIBYLL, despite its very low diffractive
probability (see figure \ref{fig:diffraction}).

We conclude our study of individual collisions by analyzing the
transverse momenta of the secondary particles.

The transverse momentum distributions, conveniently described by means
of pseudorapidity distributions, are significantly correlated with
shower observables like the lateral distribution of muons at large
distances from the core \cite{talkmalargue}, and constitute therefore
an additional source of uncertainty to be taken into account when
estimating systematic errors associated to Monte Carlo estimations of
shower observables.

\begin{figure}[tp]
\begin{center}
\includegraphics{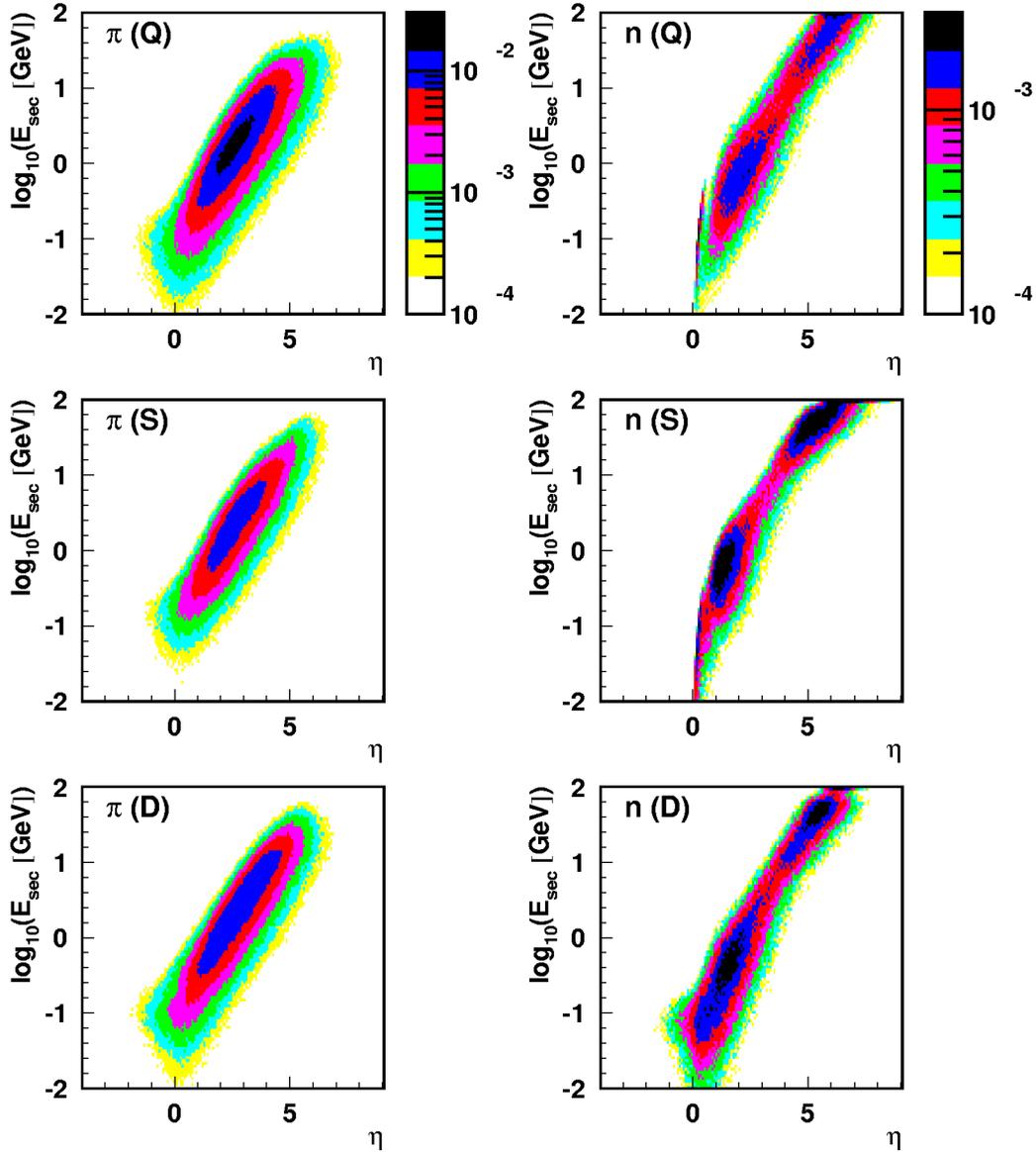}
\end{center}
\caption{Secondary pseudorapidity versus energy plots for the case of
  100 GeV proton-air collisions. The plots in the left (right) column
  correspond to pion (nucleon) distributions, for collisions generated
  with QGSJET (Q), SIBYLL (S), and DPMJET (D). The scales are the
  same at each column.}
\label{fig:pseudorap2proton100gev}
\end{figure}

\begin{figure}[tp]
\begin{center}
\includegraphics{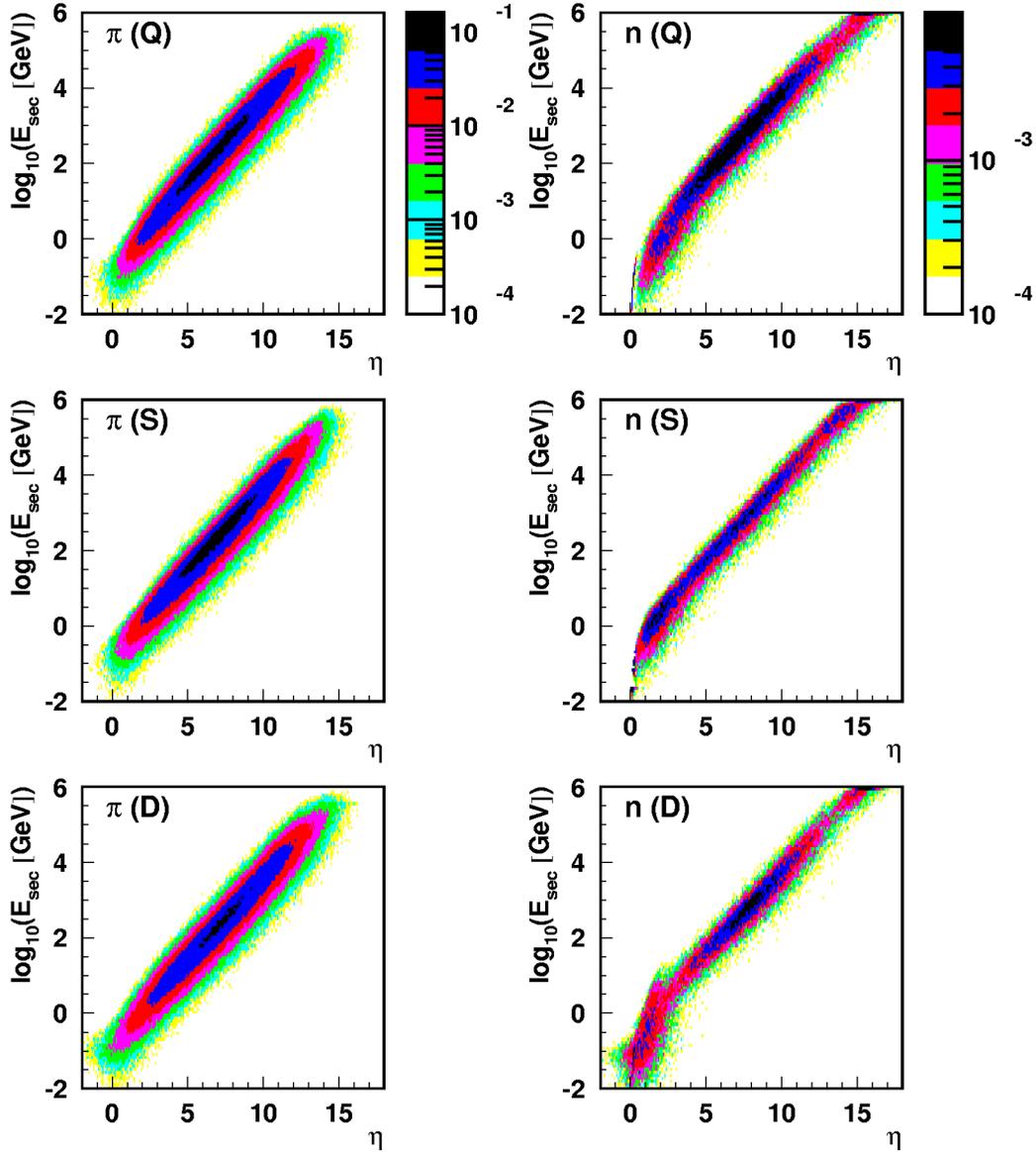}
\end{center}
\caption{Same as figure \ref{fig:pseudorap2proton100gev} but for 1 PeV
  proton-air collisions.}
\label{fig:pseudorap2proton1pev}
\end{figure}

Figures \ref{fig:pseudorap2proton100gev} and
\ref{fig:pseudorap2proton1pev} contain $\eta \times \esec$
two-dimensional distributions for proton-air collisions at 100 GeV and 1
PeV respectively.

The simplest distributions are the ones corresponding to secondary
pions. At each secondary energy the pseudorapidity $\eta$ distributes
around a central value approximately like a Gaussian, but presenting
however a longer tail towards the region of large $\eta$. The mean
pseudorapidity increases with $\log \esec$ approximately in a linear
form.

On the other hand, the distributions for nucleons are more complex,
and there are evident differences between the studied hadronic
models. In particular, both SIBYLL and QGSJET distributions do not
produce recoiling nucleons, that is, there are no $\eta < 0$ events,
as it shows up clearly in the figures. This is not the expected
behavior of the secondaries which can eventually emerge as recoiling
particles, especially for those having low energies. Notice that, on
the other side, DPMJET is capable of generating such recoiling
particles

\begin{figure}[tp]
\begin{center}
\includegraphics{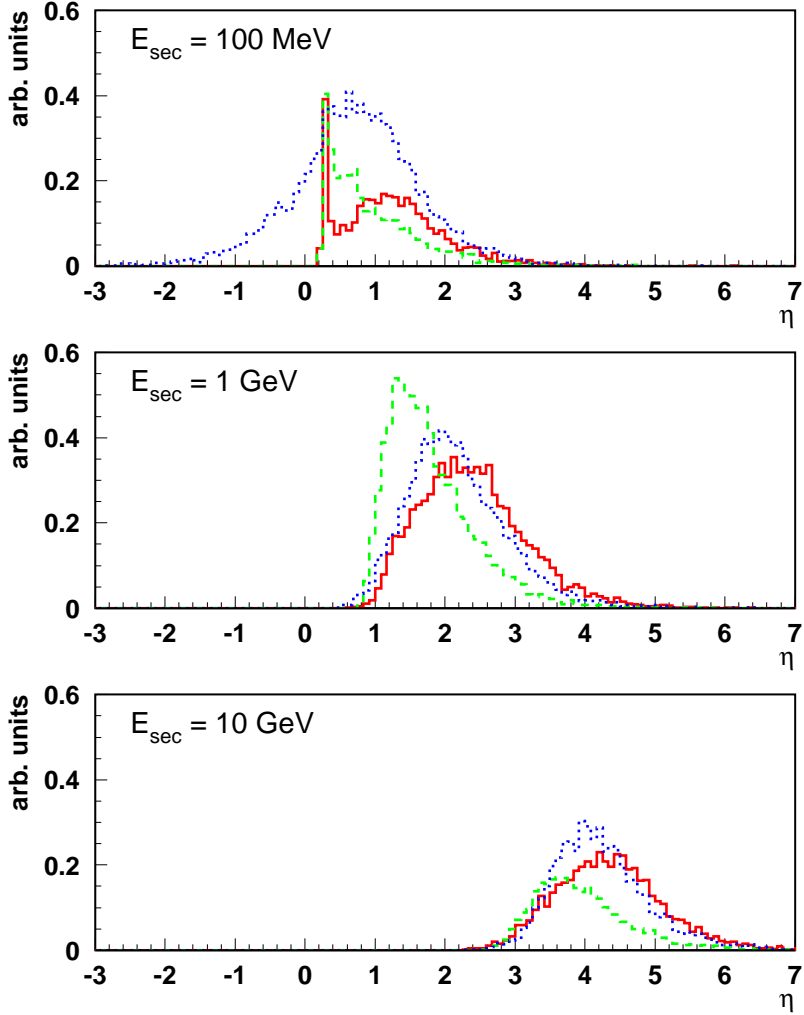}
\end{center}
\caption{Some cuts around representative secondary energies of the
  $\eta \times \esec$ distributions of figure
  \ref{fig:pseudorap2proton100gev}. The full, dashed, and dotted lines
  correspond to the QGSJET, SIBYLL, and DPMJET cases, respectively.}
\label{fig:pseudorapslicesproton100gev}
\end{figure}

In figure \ref{fig:pseudorapslicesproton100gev} we show three
representative $\eta$ distributions corresponding to three slices of the
distributions of the right column of figure
\ref{fig:pseudorap2proton100gev}, at 100 MeV, 1 GeV, and 10 GeV,
respectively.

In the 100 MeV distributions the lack of recoiling nucleons in the SIBYLL
and QGSJET cases is most evident. Notice also that in the remaining
case the distributions are similar but not completely coincident.

\section{Effect on showers observables}
\label{sec:eso}

The second step in our analysis is to study the impact that different
alternatives for modeling the diffractive hadronic interactions have
on common air shower observables.

We have used the AIRES program \cite{aires} to simulate proton and
iron induced showers with different primary energies, and using
QGSJET01 and SIBYLL 2.1 to process the high energy hadronic
interactions \footnote{AIRES uses a built-in algorithm to process the
  low energy ($E<100$ GeV) hadronic and photonuclear
  interactions. High energy photonuclear processes are simulated via
  calls to the corresponding external hadronic package. A detailed
  description of the organization and features of AIRES is provided in
  reference \cite{aires}.}.

In the previous section we have shown that the fraction of diffractive
events was one of the most outstanding differences between the
tested hadronic codes (see for example figure
\ref{fig:diffraction}). We consider therefore that it is worthwhile to
obtain quantitative estimations of the impact of the diffractive
interactions on shower observables. To this end, we have run
simulations using SIBYLL or QGSJET to process the hadronic
interactions, with two different configurations: (i) normal setting
mixing diffractive and non-diffractive events. (ii) disabling
diffractive interactions.

\subsection{Longitudinal development}

\begin{figure}[tp]
\begin{center}
\includegraphics{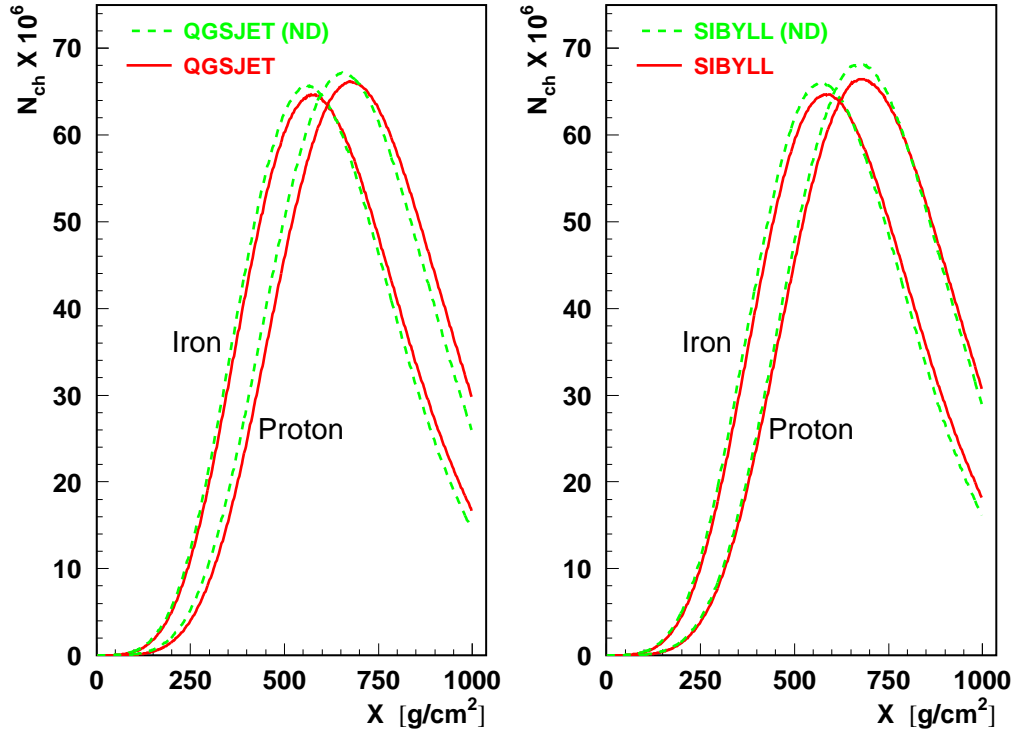}
\end{center}
\caption{Average longitudinal development of $10^{17}$ eV proton 
  and iron showers. The simulations were performed using AIRES 
  linked to QGSJET01 (a), or SIBYLL 2.1 (b).}
\label{fig:nchvsx1017}
\end{figure}

\begin{figure}[tp]
\begin{center}
\includegraphics{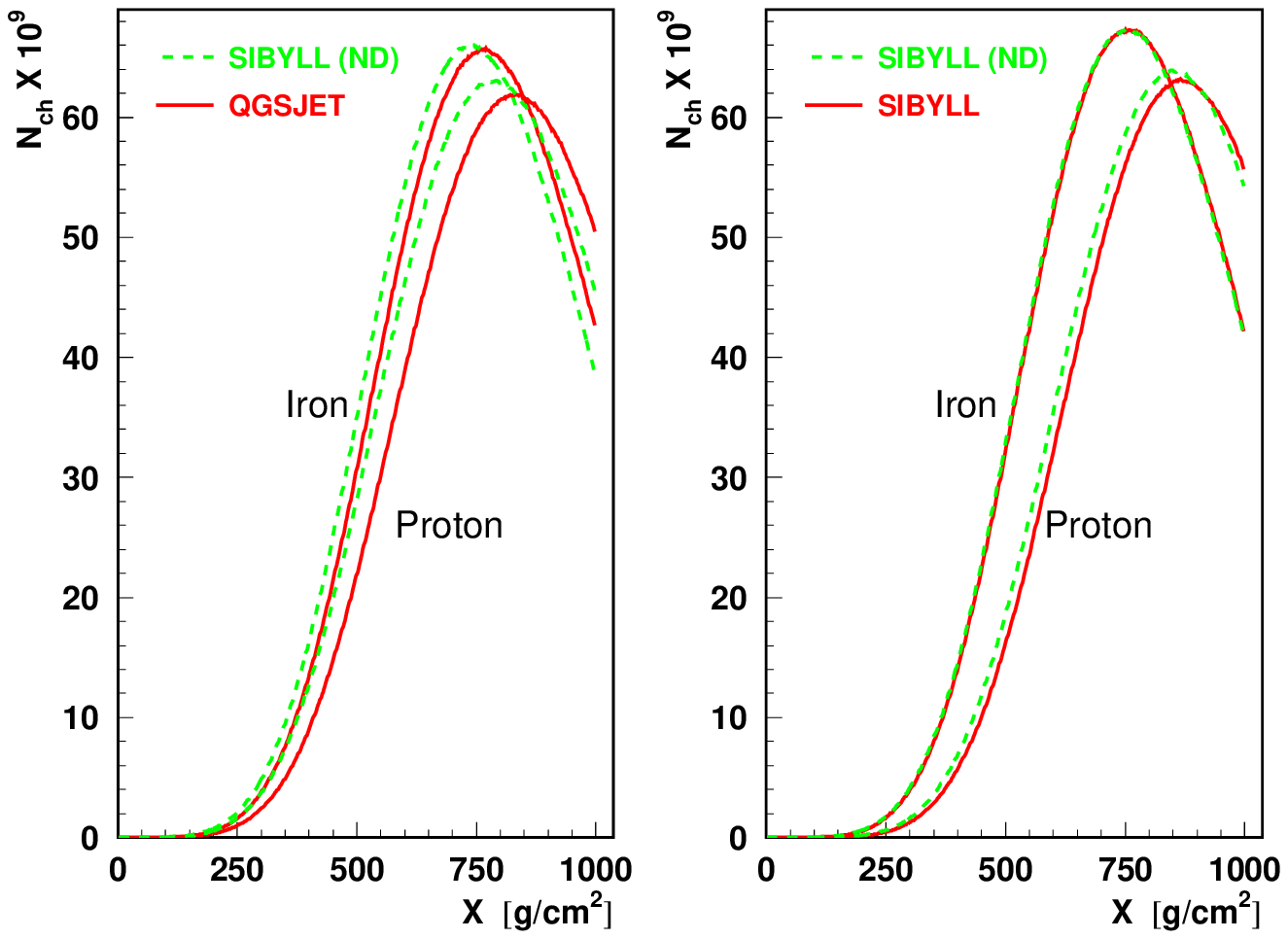}
\end{center}
\caption{Same as figure \ref{fig:nchvsx1017} but for $10^{20}$ eV
  showers.}
\label{fig:nchvsx1020}
\end{figure}

Because of their active role in energy transport, the diffractive
interactions have a direct impact on the global shower
development. This fact shows up clearly in figures
\ref{fig:nchvsx1017} and \ref{fig:nchvsx1020} where the number of
charged particles is plotted versus the atmospheric depth, in the
cases of $10^{17}$ and $10^{20}$ eV vertical proton showers,
respectively. The plots were done using data coming from simulations
performed with AIRES linked to QGSJET (a) and SIBYLL (b). As expected,
when the diffractive interactions are disabled (dotted lines), the
showers develop earlier than in the normal case. This implies a
displacement in the position of the maximum, $\xmax$, that amounts
approximately to $20$ \gcm2 for iron and $30$ \gcm2 for proton  
($10$ \gcm2 for both iron and proton) for QGSJET (SIBYLL) 
simulations.

\begin{figure}[tp]
\begin{center}
\includegraphics[width=11cm]{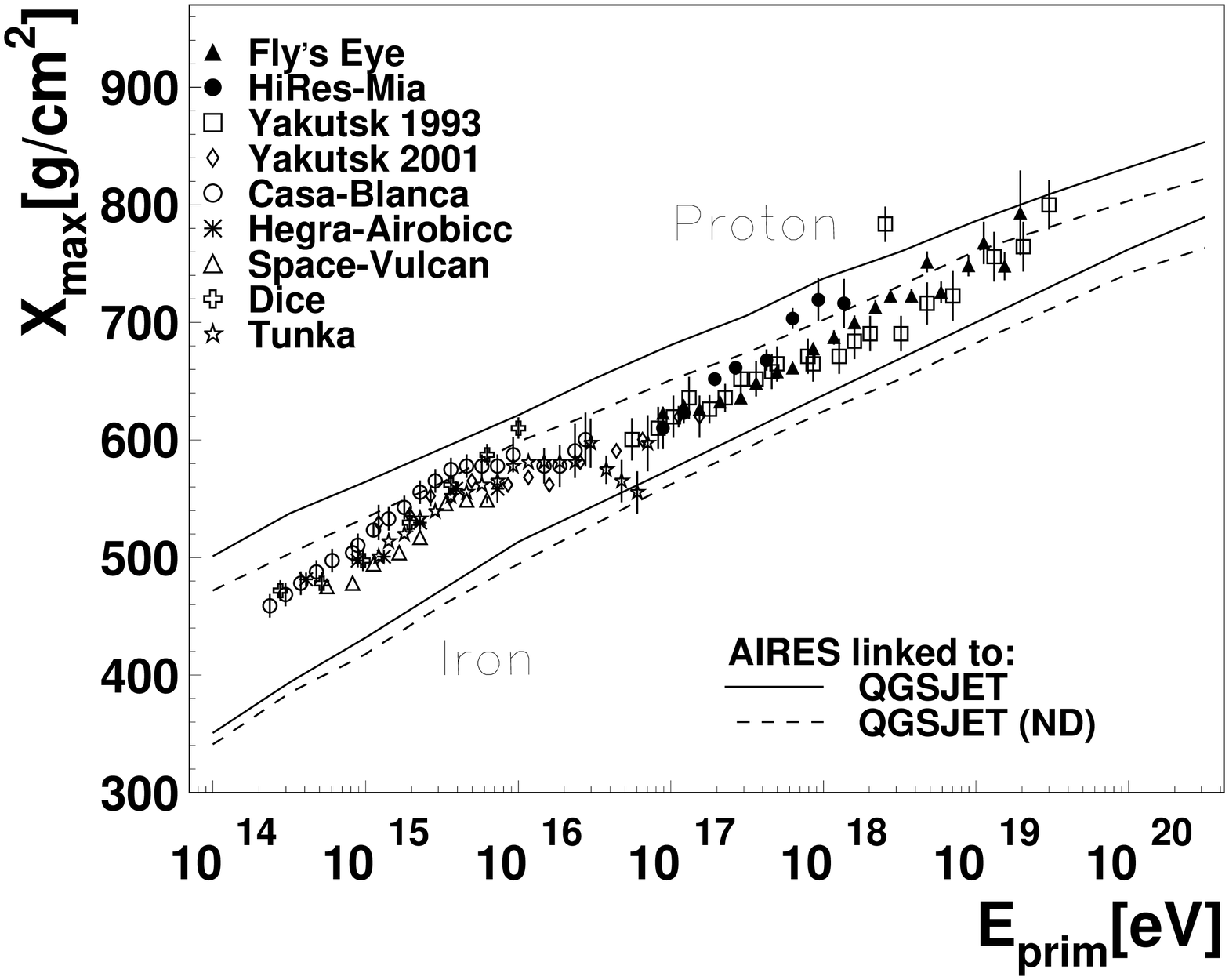}
\includegraphics[width=11cm]{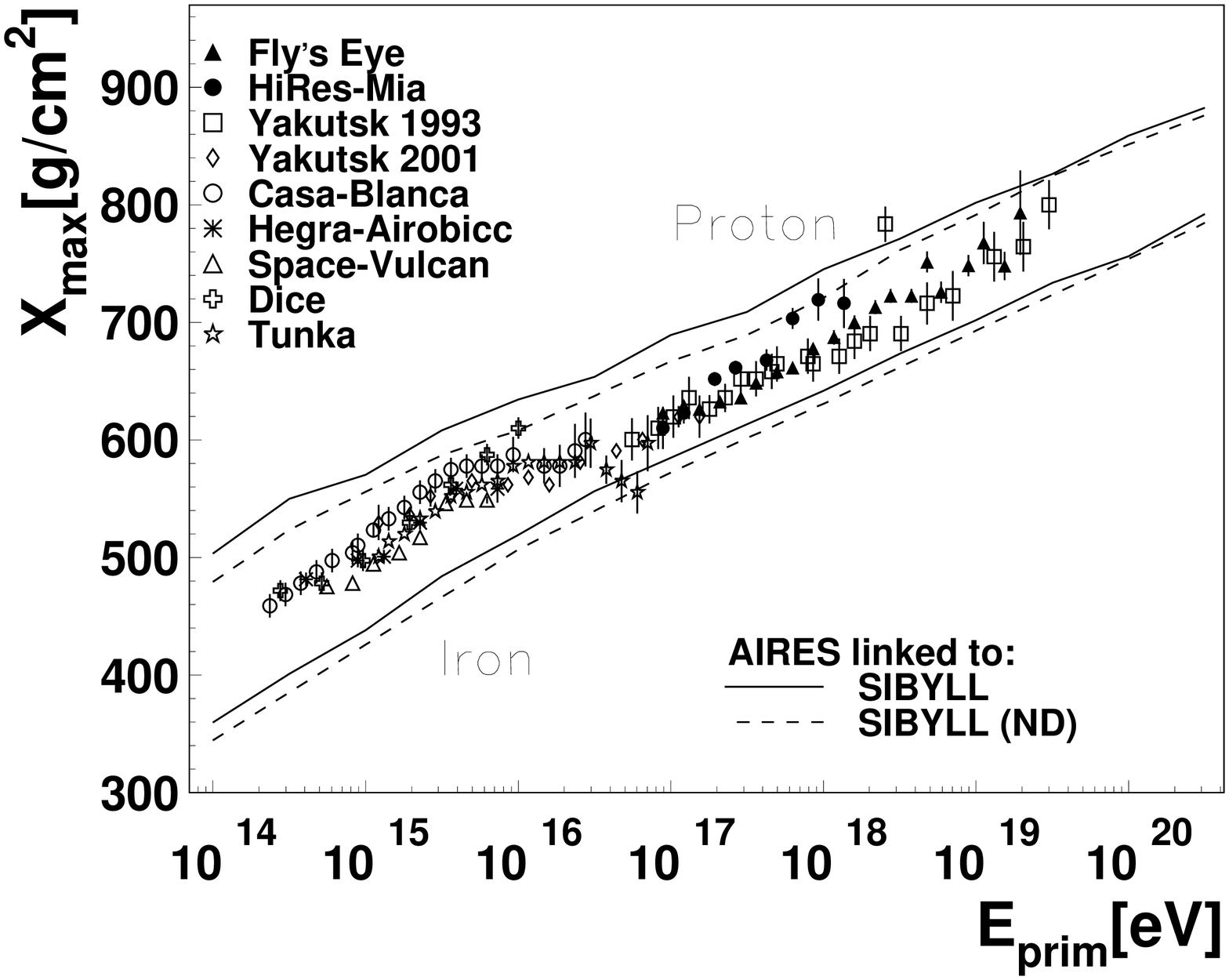}
\end{center}
\caption{Average shower maximum versus primary energy.}
\label{fig:xmaxvseprim}
\end{figure}

The shift in the position of the shower maximum, due to the
suppression of diffractive interactions is significant at all
primary energies. This is illustrated in figure
\ref{fig:xmaxvseprim} where $\xmax$ is plotted versus the primary
energy. The lines represent simulations of proton and iron showers
enabling (solid lines) or disabling (dashed lines) the diffractive
interactions. We have also plotted some available experimental data
for reference \cite{EngelReview}.

We can clearly see that in the entire covered range of primary
energies the suppression of diffractive interactions always produce a
non negligible reduction of $\xmax$. It is clear that the fully
non-diffractive simulations are unrealistic, but they are useful to
quantitatively estimate a rough upper bound of the uncertainty of
$\xmax$ that can be expected due to the uncertainties associated with
diffractive hadron-nucleus interactions, especially at the highest
energies.

Notice also that the differences for SIBYLL are generally smaller than
the corresponding ones for the QGSJET case. This is correlated with
the fact that in SIBYLL the diffractive interactions have a very small
probability, in comparison with QGSJET, as discussed in section
\ref{sec:mandi}.

\begin{figure}[tp]
\begin{center}
\includegraphics[width=6.5cm]{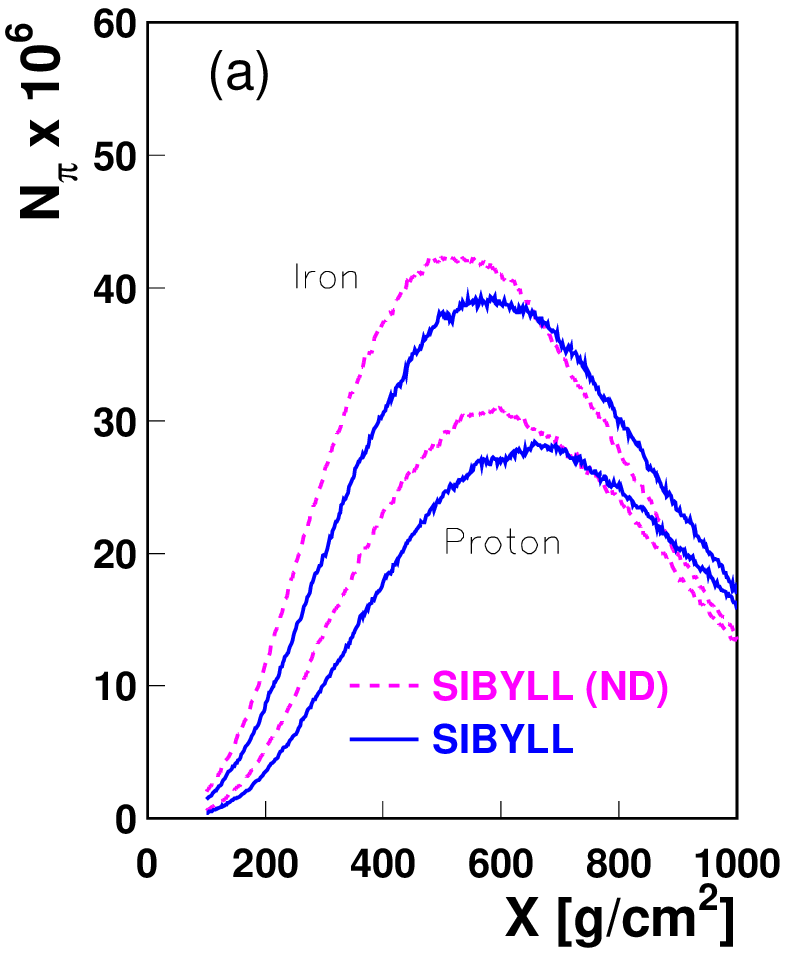}
\includegraphics[width=6.5cm]{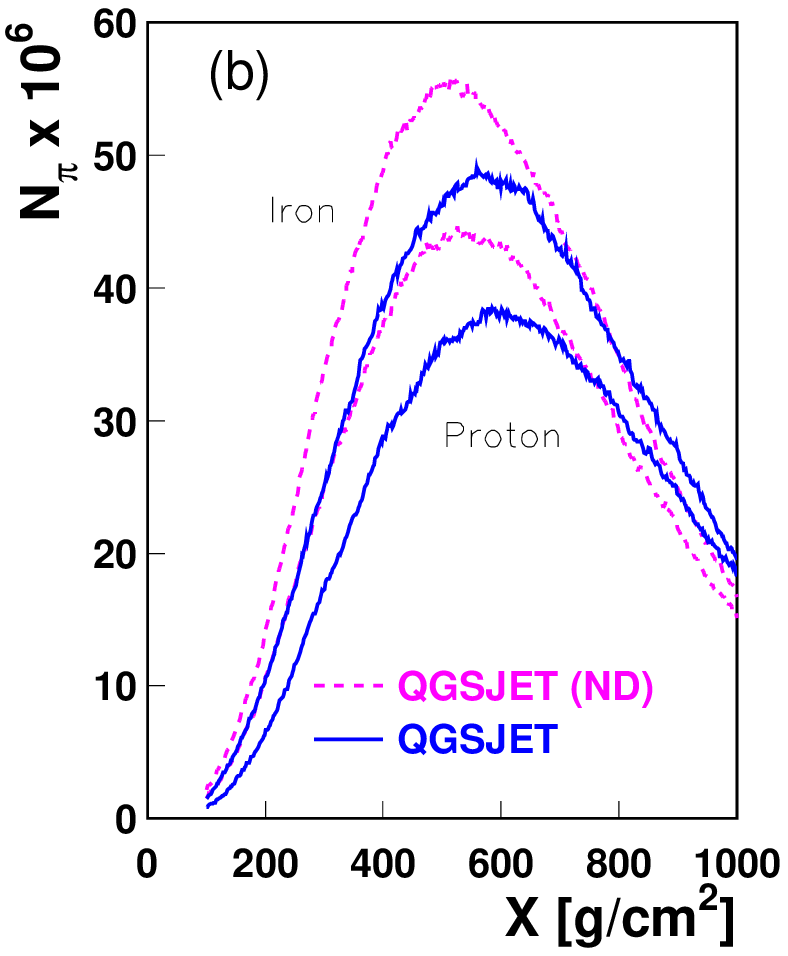}\\
\includegraphics[width=6.5cm]{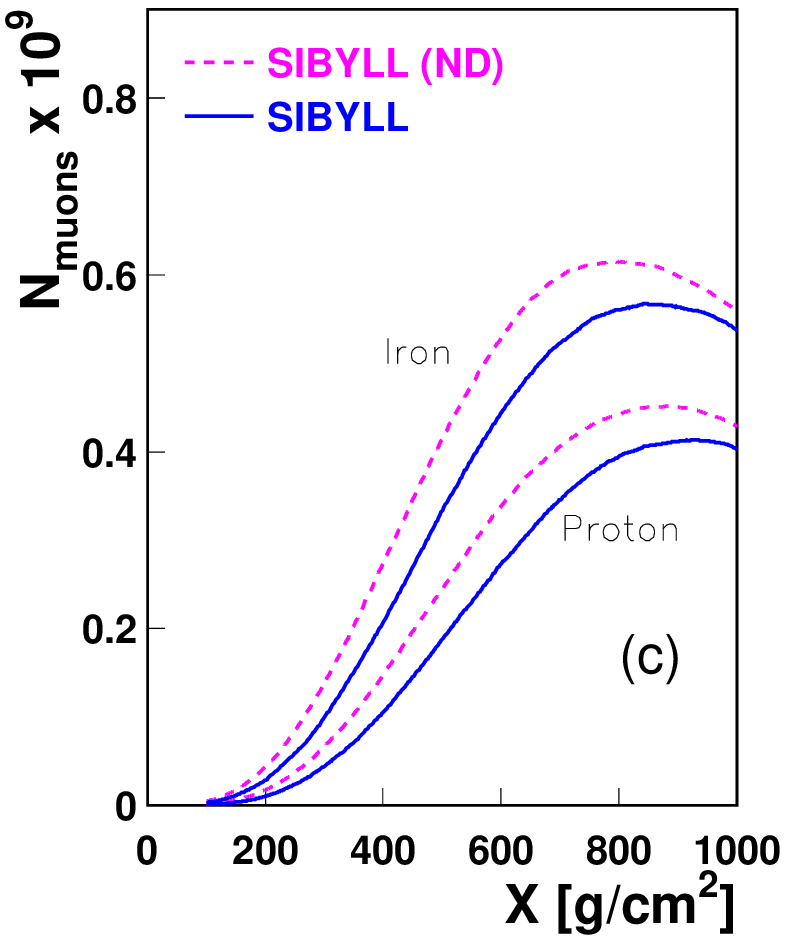}
\includegraphics[width=6.5cm]{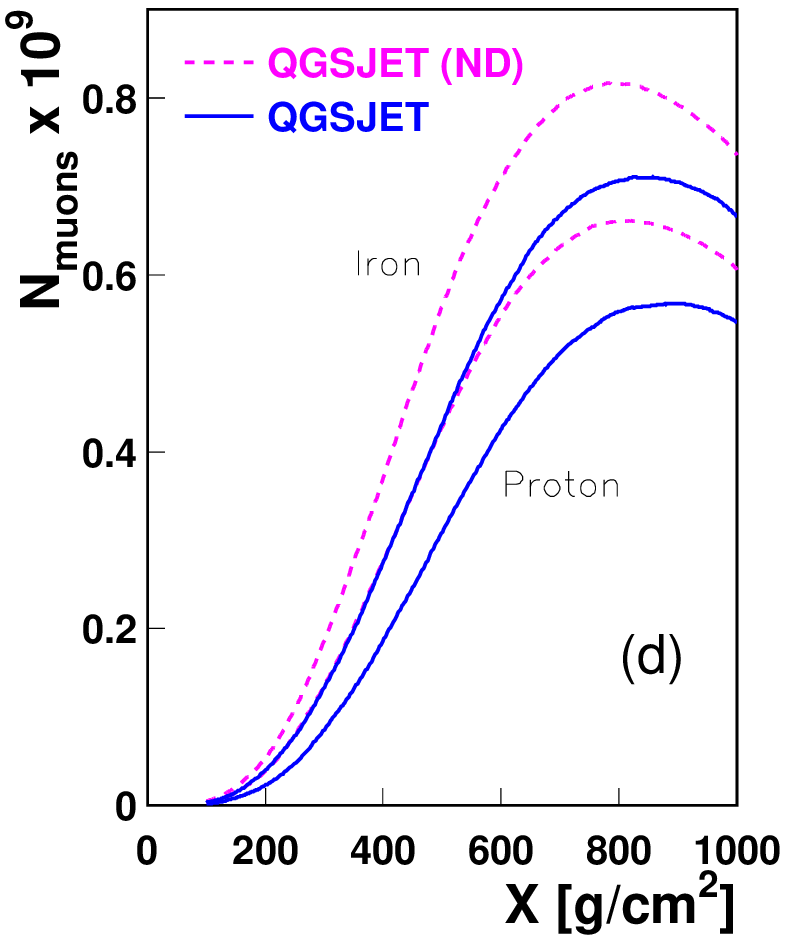}
\end{center}
\caption{Longitudinal development of charged pions and muons for showers
  initiated by $10^{20}$ eV protons.}
\label{fig:npimuvsx1020}
\end{figure}

The diffractive interactions have also a direct impact on the
development of the hadronic and muonic components of the showers. This
fact is illustrated in figure \ref{fig:npimuvsx1020} where the numbers
of pions and muons are plotted as functions of the atmospheric depth,
in the case of showers initiated by $10^{20}$ eV protons. For both
SIBYLL and QGSJET cases, the number of pions increases when the
diffractive interactions are disabled (figures \ref{fig:npimuvsx1020}a
and \ref{fig:npimuvsx1020}b). This can be clearly understood because
the bulk of the pions are produced at inelastic hadronic collisions,
whose number is enlarged when diffraction is switched off.

Muons come mainly from decays of charged pions and therefore it can be
expected that a larger number of pions leads to an increase in the
number of muons. This feature is clearly illustrated at figures
\ref{fig:npimuvsx1020}c and \ref{fig:npimuvsx1020}d, that show
that the muon numbers for the non diffractive case are larger than the
respective ones for the normal simulations.

\subsection{Lateral distributions}

The lateral distribution of particles reaching ground is a key
observable whose accurate determination is essential for the
calibration of surface array detectors
\cite{PlayaDelCarmen,DPMJET2_1,NEXUS}, like the water \v Cerenkov
tanks of the Auger experiment \cite{AugerNIM} for example. In such
detectors the primary energy is estimated from the signal measured at
a determined distance from the shower core, 1000 meters for
example. In the particular but important case of water C\v erenkov
detectors, the detected signal comes mainly from the electromagnetic
particles (gammas, electrons and positrons), and the muons. In a
simulation, the muonic part depends strongly on the characteristics of
the hadronic model used. This is due to the fact that the dominant
channel for muon production is pion decay, so the number of produced
muons is directly correlated with the number of charged pions which in
turn appear mainly during hadronic collisions.

For these reasons we have included in our study an analysis of the
correlations between the lateral distribution of muons and the
diffractive interactions that take place during shower development.

\begin{figure}[tp]
\begin{center}
\includegraphics[width=6.5cm]{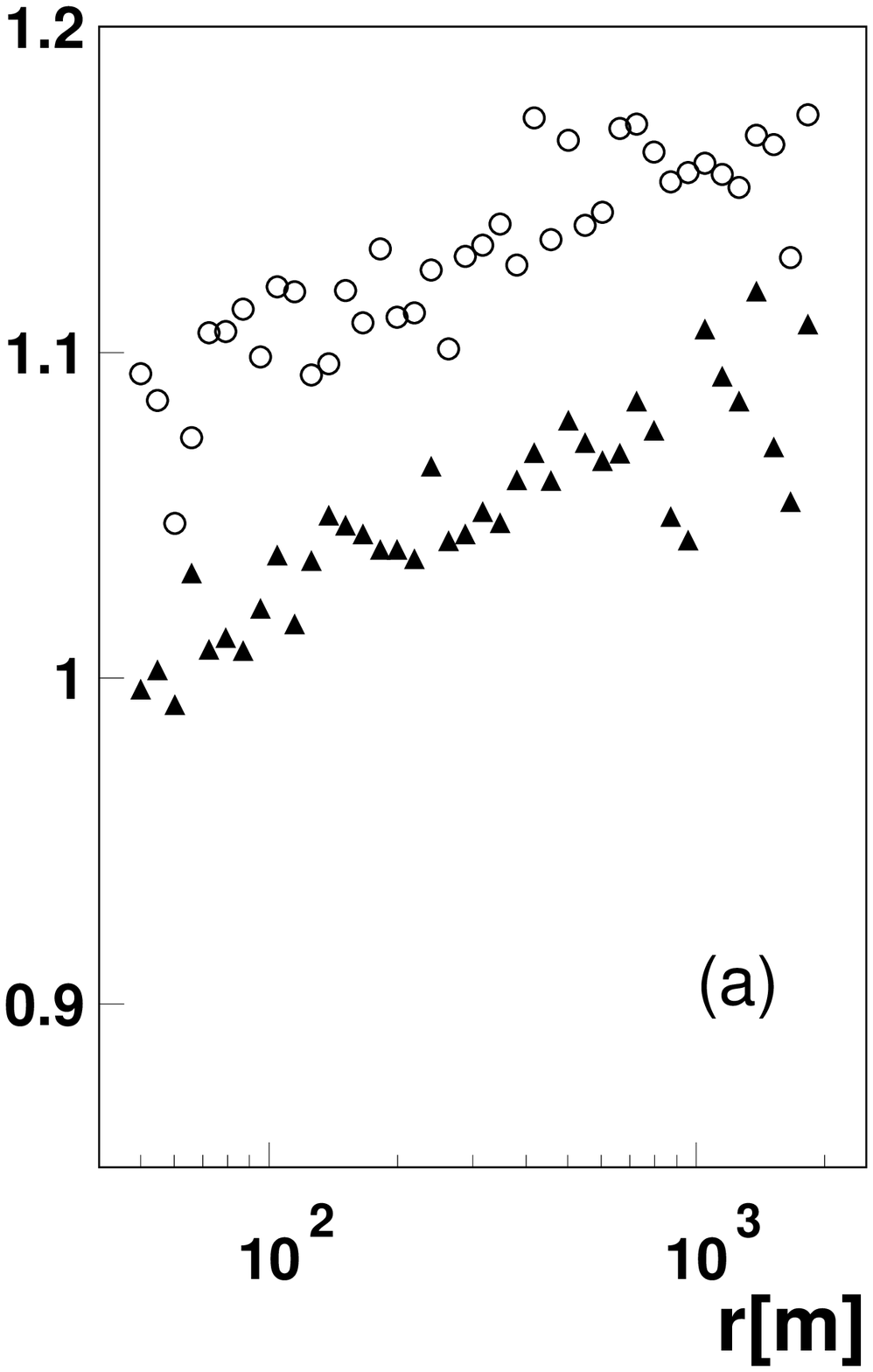}
\includegraphics[width=6.5cm]{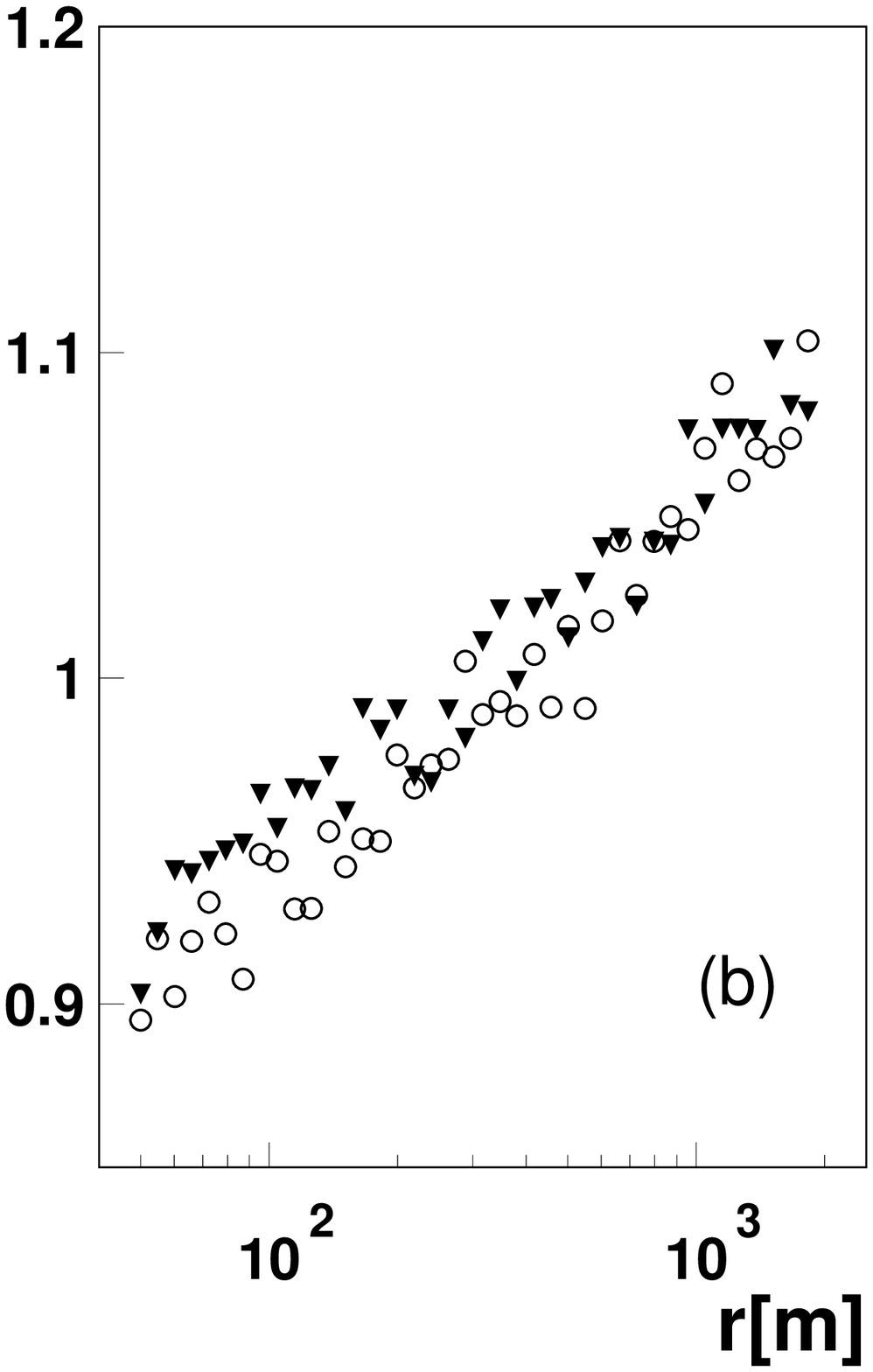}
\end{center}
\caption{Ratio between muon densities simulated disabling and enabling
diffractive interactions, plotted as a function of the distance to the
shower core. The triangles (circles) correspond to $10^{17}$ eV iron
(proton) primaries. The simulations were performed using AIRES linked
to QGSJET (a), or SIBYLL (b), and correspond to vertical showers
observed at a ground altitude of 1000 \gcm2.}
\label{fig:latdivmu}
\end{figure}

Let $\rho_\mu(r)$ be the density of muons at a given distance $r$ from
the shower core. In figure \ref{fig:latdivmu} we have plotted the
ratio $\rho^{\rm (ND)}_\mu(r)/\rho^{\rm (D)}_\mu(r)$ between muon
densities simulated disabling and enabling diffractive interactions,
as a function of $r$. When this ratio is 1, this means that enabling
or disabling the diffractive interactions that could take place during
the development of the showers, does not alter the muon density at
ground.

We can see, however, that our simulations indicate that this ratio is,
in general, different from 1, and that its behavior depends on the
hadronic model used.

In the QGSJET case (figure \ref{fig:latdivmu}a), the ratio increases
with $r$, and depends on the nature of the primary. At $r=1000$ m the
ratio is approximately 1.05 (1.15) for proton (iron) primaries.

Figure \ref{fig:latdivmu}b illustrates the SIBYLL case, characterized
by a ratio that increases with $r$. It is smaller that 1 when $r<300$
m, and greater than one otherwise. No significant composition
dependencies are present in this case, that gives 1.07 for both proton
and iron primaries at $r=1000 m$.

From both plots one can conclude that the uncertainty in the
diffractive cross sections imply an uncertainty of about 10\% in the
muon density at 1000 m from the core. Additionally, the slope of
$\rho_\mu(r)$ is found to be significantly dependent on the
diffractive cross section. This fact should be taken into account when
considering the accuracy of the primary mass estimation algorithms
that are based on measurements of the shape of the lateral
distributions.

It is also important to mention that the preceding precentages depend
on various shower parameters like primary energy, inclination, and
ground altitude. Therefore those figures should be considered only as
qualitative indicators.

As mentioned before, the electromagnetic component of the particles
reaching ground is affected in a lesser degree when switching on and
off the diffractive interactions. While a complete analysis of the
variations of measurable signals at ground detectors is beyond the
scope of this work, we can nevertheless mention that a simple analysis
leads to the conclusion that the relative variations of the total
signal are qualitatively similar to the muon densities plotted in
figure \ref{fig:latdivmu}, with an uncertainty of about 10\% at 1000 m
from the core, in the representative case of $10^{20}$ eV proton
showers inclined 30 deg.

\section{Conclusions}
\label{sec:conclu}

We have performed an extensive analysis of some of the characteristics
of commonly used hadronic collision simulation packages, namely
SIBYLL, QGSJET, and DPMJET.

The contribution of diffractive processes, as defined in section
\ref{sec:hc0}, was studied with particular detail, including an
analysis of their impact on several shower observables.

The data obtained from series of hadron-nucleus collisions simulated
using the mentioned models with identical initial conditions indicate
that there are significant differences between models for observables
such as mean multiplicity, inelasticity, fraction of pions, energy
distribution of secondaries, as well as pseudorapidity distributions
in some cases.

Such differences exist for all the primary energies that have been
studied, which include the range of moderate energies where
experimental data do exist. At such energies (of the order of 100 GeV)
the most noticeable differences correspond to the mean multiplicity
(see inset of figure \ref{fig:nsecvseproton}, and fractions of pions (see
figure \ref{fig:pifracvsegy}).

The average fractions of diffractive events detected at given primary
energies have also been studied. Our analysis puts in evidence
enormous differences between models (see figure
\ref{fig:diffraction}): for QGSJET the fraction of diffractive events
rises slowly with energy, passing from about 10\% at $E_{\rm prim}=
100$ GeV to 13\% at $E_{\rm prim}= 10^{20}$ eV. On the other hand, for
SIBYLL this fraction diminishes with energy, from about 12\% at
$E_{\rm prim}= 200$ GeV, down to about 1\% at $E_{\rm prim}= 10^{20}$
eV. DPMJET presents an ``intermediate'' behavior as illustrated in
figure \ref{fig:diffraction}.

The fraction of diffractive events is directly related to the
diffractive to total cross section ratio. The data plotted in figure
\ref{fig:diffraction} seem to indicate that the different models have
significantly different ways of extrapolating those cross sections for
the case of extremely large energies, while presenting qualitatively
similar values at primary energies around 1 TeV.

Our study is completed with an analysis of the impact of the
diffractive events on common shower observables. We have run several
shower simulations using SIBYLL and QGSJET, enabling or disabling the
diffractive events, with the main purpose of extracting conclusions
about the maximum impact that the uncertainty in the diffractive
probability can have on the considered observables.

We have found a moderate but not negligible impact for both the
position of the shower maximum, $X_{\rm max}$, and the lateral
distribution of muons, $\rho_{\mu}(r)$. In  this last case, it is
remarkable the change of slope detected when changing the probability
of diffractive events.

It is worthwhile mentioning that in the case of $\xmax$, the detected
differences are of the order of about 2\%, and this figure is of the
same order of magnitude than other uncertainties in $\xmax$ connected
with the hadronic model, that have been already reported in a previous
work \cite{augersimap}.

The detected discrepancies between models call for further studies,
both theoretical and experimental. In this last case, the possibility
of measuring fractions of diffractive events and multiplicities at
primary energies larger than 1 TeV will certainly help to improve the
constraints needed to validate a given model.

\acknowledgments

We are indebted to T. Gaisser, R. Engel, H. Fanchiotti and R. Sassot
for useful discussions and comments. This work was partially supported
by ANPCyT and Fundaci\'on Antorchas of Argentina, and SUPERA-SEP and
CONACYT of Mexico.

\def\journal#1#2#3#4{{\em #1,\/} {\bf #2}, #3 (#4)}

\end{document}